\documentclass{cimento}

\usepackage{url}
\title{
Data-driven approximations to the Hadronic Light-by-Light scattering contribution to the muon (g-2)}
\author{P.~Masjuan\from{ins:x} \from{ins:y}\ETC\,
and P.~Roig\from{ins:z}}
\instlist{\inst{ins:x} 
Grup de Física Teòrica, Departament de Física, Universitat Autònoma de Barcelona, 08193
Bellaterra (Barcelona), Spain.
\inst{ins:y}
Institut de Física d’Altes Energies (IFAE) and The Barcelona Institute of Science and
Technology (BIST), Campus UAB, 08193 Bellaterra (Barcelona), Spain.
  \inst{ins:z} 
 Departamento de Física, Centro de Investigación y de Estudios Avanzados del Instituto
Politécnico Nacional, Apdo. Postal 14-740,07000 Ciudad de México, México. 
  }
\begin{document}

\maketitle

\begin{abstract}
We review recent progress on the numerical determination of the Hadronic Light-by-Light contribution to the anomalous magnetic moment of the muon%, discussing the role of experimental data on its accuracy
. We advocate for a slight increase of the White Paper number for its Standard Model prediction, to $(102\pm17)\times10^{-11}$, accounting for a revised contribution from axial-vector mesons and short-distance constraints. This $\sim10\%$ larger result seems to be supported by the most recent lattice QCD evaluations.
\end{abstract}

\section{Why it matters?}
The Standard Model (SM) uncertainty on the Muon g-2 ($2a_\mu=g_\mu-2$) is dominated by the hadronic vacuum polarization (HVP) piece, amounting to $4.0\times10^{-10}$ (for an overall error of $4.3\times10^{-10}$) \cite{Aoyama:2020ynm}~\footnote{These and the following numbers are quoted -unless otherwise stated- from the Whitepaper of the Muon g-2 Theory Initiative \cite{Aoyama:2020ynm} (WP), a collaboration which has been aiming for a community consensus value of the Standard Model prediction of the muon g-2, see \url{https://muon-gm2-theory.illinois.edu/}.}. This is contributed very mildly by the error of the Hadronic light-by-light (HLbL) scattering part, $1.9\times10^{-10}$, that we will discuss here~\footnote{See talks focusing on diverse aspects of the HVP contribution by Matthia Bruno, Christoph Redmer, Francesca de Mori, Álex Miranda, Camilo Rojas and David Díaz Calderón.}.
Clearly, the most urgent thing is to clarify the discrepancy between the data-driven results \cite{Aoyama:2020ynm,Colangelo:2018mtw,Hoferichter:2019mqg,Davier:2019can,Keshavarzi:2019abf} and the competitive lattice QCD evaluation, by the BMW collaboration \cite{Borsanyi:2020mff}, of $a_\mu^{\rm HVP}$. To this end, several approaches have been developed, exploiting the so-called windows in Euclidean time \cite{RBC:2018dos,Lehner:2020crt,Giusti:2021dvd,Colangelo:2022jxc,Colangelo:2022vok,Ce:2022kxy,ExtendedTwistedMass:2022jpw,FermilabLattice:2022izv,Blum:2023qou,Masjuan:2023qsp,Davier:2023cyp}. Ref. \cite{Masjuan:2023qsp} (based on the isospin-breaking corrections computed in Ref.~\cite{Miranda:2020wdg}), points to nice agreement between data-driven predictions using $\tau^-\to\pi^-\pi^0\nu_\tau$ data \cite{OPAL:1998rrm,CLEO:1999dln,ALEPH:2005qgp,Belle:2008xpe} (instead of $e^+e^-\to\pi^+\pi^-$ measurements) with lattice QCD evaluations. The barely acceptable discrepancy between KLOE \cite{KLOE:2004lnj,KLOE:2008fmq,KLOE:2010qei,KLOE:2012anl,KLOE-2:2017fda} and BaBar \cite{BaBar:2009wpw,BaBar:2012bdw} $e^+e^-\to\pi^+\pi^-$ data has been aggravated by the new CMD-3 measurement \cite{CMD-3:2023alj}, being this puzzle still not understood (see also e.g. the measurements~\cite{CMD-2:2006gxt,BESIII:2015equ,SND:2020nwa}). Amid this conundrum, halving the error of the SM prediction for $a_\mu^{HLbL}$ \cite{Aoyama:2020ynm} is still necessary, according to the final precision that the Fermilab experiment will achieve measuring $a_\mu$, but not the top priority.

On the contrary, the experimental situation seems crystal-clear: FNAL measurements \cite{Muong-2:2021ojo,Muong-2:2023cdq} are extremely consistent with the BNL outcome \cite{Muong-2:2006rrc} and their joint picture is fully convincing, yielding
\begin{equation}
a_\mu^\mathrm{Exp}=116592059(22)\times10^{-11}\,.\end{equation}
 This situation further enhances the pressing need for theoretical progress.

\section{Why such a large error for $a_\mu^{\rm HLbL}$?}
The outsider may wonder why the uncertainty of the $a_\mu^{\rm{HLbL}}$ is $\mathcal{O}(20\%)$, while that of the $a_\mu^{\mathrm{HVP}}$ is only $\mathcal{O}(0.6)\%$. This much better precision stems from its calculation via a single dispersive integral that is related to the  accurately measured  $\sigma(e^+e^-\to\mathrm{hadrons})$ \cite{Brodsky:1967sr} plus a mild contribution from perturbative QCD. On the contrary, a data-driven approach to $a_\mu^{\mathrm{HLbL}}$ is very much complicated by the additional loop and multi-scale nature of the problem. Despite enormous advances towards a fully-dispersive computation of $a_\mu^{\rm HLbL}$ \cite{Colangelo:2014dfa,Colangelo:2014pva,Pauk:2014rfa,Colangelo:2015ama,Colangelo:2017qdm,Colangelo:2017fiz}, a completely dispersive evaluation is not feasible yet. This framework provided a rationale for the historical arrangement of the main contributions (starting from the dominance of the pseudoscalar-pole cuts~\cite{Knecht:2001qf}) and could in principle be used up to arbitrary complex multiparticle ones.

\section{Contributions}
Amazingly, the whole $a_\mu^{\rm HLbL}$ is basically saturated by the contribution from the lowest-multiplicity cut (even more so because of the approximate cancellations among the other contributions), corresponding to the lightest pseudoscalar ($\pi^0,\,\eta,\,\eta'$) poles, yet it could be related to a combined chiral and large-$N_C$ expansion \cite{deRafael:1993za}. This can be computed straightforwardly \cite{Knecht:2001qf} knowing the corresponding pseudoscalar transition form factors (TFFs) as functions of both photons virtuality. See Christoph Redmer's talk on the precious experimental input to these (and others required for $a_\mu^{\rm HLbL}$) TFFs. In addition, there are some theoretical properties constraining these TFFs, like the chiral limit, the singly and doubly virtual asymptotic limits predicted by QCD, analyticity and unitarity ...
The dispersive evaluation \cite{Hoferichter:2018dmo,Hoferichter:2018kwz} yields a very precise result for the $\pi^0$ contribution
\begin{equation}
a_\mu^{\pi^0,{\rm HLbL}}=\left(63.0^{+2.7}_{-2.1}\right)\times10^{-11}\,,
\end{equation}
confirming the rational approximants' determination \cite{Masjuan:2017tvw}
\begin{equation}
a_\mu^{\pi^0,{\rm HLbL}}=\left(63.6\pm2.7\right)\times10^{-11}\,.
\end{equation}
 These results are also supported by e.g. Dyson-Schwinger eqs. evaluations, yielding $a_\mu^{\pi^0,{\rm HLbL}}=\left(62.6\pm1.3\right)\times10^{-11}$ \cite{Eichmann:2019tjk}, and $a_\mu^{\pi^0,{\rm HLbL}}=\left(61.4\pm2.1\right)\times10^{-11}$ \cite{Raya:2019dnh} and by holographic QCD results \cite{Cappiello:2019hwh,Leutgeb:2021mpu,Leutgeb:2022lqw} (see, however, \cite{Colangelo:2023een}) and chiral Lagrangians including resonances \cite{Roig:2014uja,Guevara:2018rhj}.
For the $\eta^{(')}$ contributions there is no dispersive computation yet. The rational approximants' calculation \cite{Escribano:2013kba,Escribano:2015nra,Escribano:2015yup,Masjuan:2017tvw} yields
\begin{equation}
a_\mu^{\eta,{\rm HLbL}}=\left(16.3\pm1.4\right)\times10^{-11}\,,\quad
a_\mu^{\eta,{\rm HLbL}}=\left(14.5\pm1.9\right)\times10^{-11}\,,
\end{equation}
which are the reference values for this contribution. Again, they are supported by the different approaches mentioned before where, in particular, Dyson-Schwinger eqs. results in $a_\mu^{\eta,{\rm HLbL}}=\left(15.8\pm1.2\right)\times10^{-11}$,  $a_\mu^{\eta',{\rm HLbL}}=\left(14.7\pm1.9\right)\times10^{-11}$ \cite{Eichmann:2019tjk} and $a_\mu^{\eta,{\rm HLbL}}=\left(13.3\pm0.9\right)\times10^{-11}$,  $a_\mu^{\eta',{\rm HLbL}}=\left(13.6\pm0.8\right)\times10^{-11}$ \cite{Raya:2019dnh}, respectively.
From the dispersive and rational approximants calculations, the WP quotes
\begin{equation}\label{eq.Ppoles}
  a_\mu^{\pi^0+\eta+\eta',{\rm HLbL}}=\left(93.8^{+4.0}_{-3.6}\right)\cdot10^{-11}\,,
\end{equation}
still to be considered the data-driven SM prediction for this leading contribution to ${\rm HLbL}$, coming from the lightest pseudoscalar poles.

The very well-known pseudoscalar electromagnetic form factors are the key objects to determine their box contributions to $a_\mu^{{\rm HLbL}}$. The dispersive result for the $\pi$ case
\begin{equation}\label{eq.pibox}
a_\mu^{\pi-box,{\rm HLbL}}=-(15.9\pm0.2)\times10^{-11},
\end{equation}
was later on confirmed by Schwinger-Dyson evaluations $a_\mu^{\pi-box,{\rm HLbL}}=-(15.7\pm0.4)\times10^{-11}$ \cite{Eichmann:2019tjk}, and $ a_\mu^{\pi-box,{\rm HLbL}}=-(15.6\pm0.2)\times10^{-11}$ \cite{Miramontes:2021exi}.
For the Kaon case, the early evaluation of ref.~
\cite{Eichmann:2019bqf}, $a_\mu^{K-box,{\rm HLbL}}=-(0.46\pm0.02)\times10^{-11}$ was slightly revised within Dyson-Schwinger and then also using a dispersive framework \cite{Stamen:2022uqh}, both agreeing on
\begin{equation}\label{eq.Kbox}
a_\mu^{K-box,{\rm HLbL}}=-(0.48\pm0.02)\times10^{-11}\,.
\end{equation}
The SM prediction comes from eqs.~(\ref{eq.pibox}) and (\ref{eq.Kbox}), still coinciding with the WP number \cite{Aoyama:2020ynm}
\begin{equation}\label{eq.Pboxes}
a_\mu^{(\pi/K)-box,{\rm HLbL}}=-(16.4\pm0.2)\times10^{-11}\,.
\end{equation}

Now we turn to another contribution coming from two-particle cuts, that associated to pseudoscalars rescattering. For the pions case, the dispersive evaluation \cite{Colangelo:2017qdm,Colangelo:2017fiz} is quite precise for the contribution associated to the $\pi$-pole left-hand cut (LHC):
\begin{equation}
a_{\mu,J=0}^{\pi\pi,\pi-pole LHC}=-(8\pm1)\times10^{-11}\,,
\end{equation}
where contributions from $D-$ and higher orders partial waves were covered by the uncertainty. This agrees with other evaluations \cite{Pauk:2014rta,Knecht:2018sci,Cappiello:2021vzi,Danilkin:2021icn} that include additional scalar contributions, converging to \cite{Danilkin:2021icn}
\begin{equation}\label{eq.Scalars}
    a_\mu^{Scalars}=-(9\pm1)\times10^{-11}\,,
\end{equation}
again in accord with the WP \cite{Aoyama:2020ynm}. Similarly, the tensors contribution \cite{Danilkin:2016hnh}
\begin{equation}\label{eq.Tensors}
    a_\mu^{Tensors}=-(0.9\pm0.1)\times10^{-11}\,,
\end{equation}
is unchanged with respect to Ref.~\cite{Aoyama:2020ynm}.

The part which has been evolving less trivially since 2020 corresponds to the axial-vector contributions, which should be regarded together with the remaining perturbative QCD constraints.

Melnikov and Vainshtein \cite{Melnikov:2003xd} put forward that pseudoscalar poles alone cannot satisfy short-distance QCD restrictions and emphasized the importance of axial-vectors to fulfil this requirement. Modern studies coincide in smaller values for these contributions than initially advocated.

Ref.~\cite{Roig:2019reh} clarified ambiguities about bases arising because of axials off-shellness and, together with ref.~\cite{Masjuan:2020jsf}, emphasized the relationship between short-distance, axial anomaly constraints, and the axial contributions (with possible relevant role of pseudoscalar resonances, see also \cite{Raya:2019dnh}), a hot topic since then. Refs.~\cite{Pauk:2014rfa,Jegerlehner:2017gek,Roig:2019reh} gave rise to the WP number \cite{Aoyama:2020ynm}
\begin{equation}
a_\mu^{Axials}=(6\pm6)\times10^{-11}\,.
\end{equation}
This was accompanied by the estimation of the contribution from light-quark loops and remaining QCD short-distance constraints ($SDCs$)~\cite{Bijnens:2019ghy,Colangelo:2019uex,Colangelo:2019lpu}
\begin{equation}
a_\mu^{u/d/s-loops+SDCs}=(15\pm10)\times10^{-11}\,.
\end{equation}
Given their correlation, these two contributions were combined with errors added linearly (uncertainties are combined quadratically, unless otherwise stated) to~\cite{Aoyama:2020ynm}
\begin{equation}\label{eq.AxialsSDCsWP}
a_\mu^{axials+SDCs}=(21\pm16)\times10^{-11}\,.
\end{equation}
Finally, the $c$-quark contribution (with uncertainty to be added linearly to the Eq.~(\ref{eq.AxialsSDCsWP})) is \cite{Raya:2019dnh,Masjuan:2012qn,Bijnens:2019ghy,Colangelo:2019uex,Colangelo:2019lpu}
\begin{equation}\label{eq.cloop}
a_\mu^{c-loop}=(3\pm1)\times10^{-11}\,.
\end{equation}
The leading-order $a_\mu^{\rm HLbL}$ contributions is obtained from Eqs.~(\ref{eq.Ppoles}), (\ref{eq.Pboxes}), (\ref{eq.Scalars}), (\ref{eq.Tensors}),   Eqs.~(\ref{eq.AxialsSDCsWP}), and (\ref{eq.cloop}), yielding
\begin{equation}\label{eq.HLbLLOWP}
a_\mu^{{\rm HLbL},LO}=(92\pm19)\times10^{-11}\,.
\end{equation}
Progress since the WP on axials and/or SDCs has improved the understanding of the regime where all photon virtualities are large, and when one of them is much smaller than the other two \cite{Hoferichter:2020lap, Knecht:2020xyr, Masjuan:2020jsf, Ludtke:2020moa, Bijnens:2020xnl, Bijnens:2021jqo, Zanke:2021wiq, Colangelo:2021nkr, Leutgeb:2021bpo,Miranda:2021lhb, Bijnens:2022itw, Radzhabov:2023odj, Ludtke:2023hvz, Hoferichter:2023tgp}. However, %results from holographic QCD and Regge models 
different model calculations considering axial-vector mesons and SDCs \cite{Leutgeb:2019gbz,Cappiello:2019hwh,Masjuan:2020jsf, Leutgeb:2021bpo,Leutgeb:2021mpu,Leutgeb:2022lqw} suggest a shift in the central value around 
\begin{equation}\label{eq.AxialSDCs}
a_\mu^{axials+SDCs}=(31\pm10)\times10^{-11}\,,
\end{equation}
larger than previously estimated, (\ref{eq.AxialsSDCsWP}), but compatible within errors. 
%are challenging the number (\ref{eq.AxialsSDCsWP}), which instead 
%From the results in \cite{Masjuan:2020jsf, Leutgeb:2021bpo} one would get 
Using Eq.~(\ref{eq.AxialSDCs}), the overall contribution would then be
\begin{equation}
a_\mu^{HLbL,LO}=(102\pm17)\times10^{-11}\,, 
\end{equation}
which is closer to the latest lattice QCD evaluations by the Mainz \cite{Chao:2021tvp} ($(109.6\pm15.9)\times10^{-11}$) and RBC/UKQCD \cite{Blum:2023vlm} ($(124.7\pm14.9)\times10^{-11}$) collaborations (to be compared to $(78.7\pm35.4)\times10^{-11}$ \cite{Blum:2019ugy} by RBC, used in the WP). At $NLO$ \cite{Colangelo:2014qya} the central value and its uncertainty are increased by only $(2\pm1)\times10^{-11}$.

These observations evince that a better understanding of the role of axial vector mesons and the intermediate energy region is an important step towards a more precise and reliable estimate for the HLbL contribution. Progress in this direction continues~\cite{Hoferichter:2020lap, Knecht:2020xyr, Masjuan:2020jsf, Ludtke:2020moa, Bijnens:2020xnl, Bijnens:2021jqo, Zanke:2021wiq, Colangelo:2021nkr, Leutgeb:2021bpo,Miranda:2021lhb, Bijnens:2022itw, Radzhabov:2023odj, Ludtke:2023hvz, Hoferichter:2023tgp}.

\section{Conclusions}
\begin{itemize}
\item The WP number, $a_\mu^{{\rm HLbL},LO}=(92\pm19)\times10^{-11}$ \cite{Aoyama:2020ynm}, still stands as the data-driven SM prediction for $a_\mu^{{\rm HLbL},LO}$.
\item The dominant uncertainty comes from short-distance + axial contributions (correlated uncertainties), with improved understanding since the WP, where still work needs to be done. This may shift the SM prediction slightly, to $a_\mu^{{\rm HLbL},LO}=(102\pm17)\times10^{-11}$.
\item Measurement of di-photon resonance couplings (particularly for axials) would be very helpful.
\item Lattice QCD has just reached a comparable  uncertainty to the data-driven determinations of this piece, thereby reducing the uncertainty through their combination to $\leq10\times10^{-11}$, in agreement with the sought accuracy by the time of the final publication of the $a_\mu$ measurement by the FNAL experiment. So the ball is on ${\rm HVP}$'s court.
\end{itemize}
%\begin{figure}
%\includegraphics{foo}     % includes figure foo.eps
%\caption{Description of the figure.}
%\end{figure}

\acknowledgments
The authors acknowledge the organizers of this excellent conference. P.~R.~ was supported by Conacyt and Cinvestav. We acknowledge Pablo S\'anchez-Puertas for nice collaborations and his insightful comments on this manuscript.

\end{document}